\documentclass{ptptex}

\usepackage{graphicx,subfigure}
\usepackage{psfrag}
\usepackage{wrapft}

\def\nn{\nonumber}

\title{Fractional flux periodicity in tori made of square lattice}

\author{
K. Sasaki\thanks{E-mail address: sasaken@imr.tohoku.ac.jp}, Y. Kawazoe
and R. Saito$^1$ 
}

\inst{
Institute for Materials Research, Tohoku University, 
Sendai 980-8577, Japan \\
$^1$Department of Physics, Tohoku University and CREST, JST,
Sendai 980-8578, Japan 
}

\abst{
 This paper describes a study on fractional flux periodicity of the
 ground state in planar systems made from a square lattice whose
 boundary is compacted into a torus.
 It is pointed out that, in the long length and large diameter limit of
 a torus, the ground-state energy and persistent currents show a
 fractional period of the fundamental unit of magnetic flux depending on
 the twist around the torus axis.
 We discuss the possible relationship between the twist and genus of a
 torus, and comment on fractional flux periodicity in a cylinder.
 }

\begin{document} 
\maketitle

\section{Introduction}

The fact that a single electron wave function has a fundamental unit of
magnetic flux $\Phi_0 = 2\pi/e$~\footnote{The electron charge is denoted
by $-e$. We will use the units of $\hbar =c =1$.} is known to be an
essential consequence of quantum mechanics and is manifest in the
Aharonov-Bohm effect~\cite{AB}.
The solid state properties of materials are governed by many electrons,
and each constituent has the $\Phi_0$ periodicity. 
However, the fundamental flux period of a material is not always
$\Phi_0$.
For instance, superconductors exhibit a period of $\Phi_0/2$,
which can be understood by charge doubling due to the Cooper pair
formation.
This must be a trivial fact if one suppose that a fundamental particle
(or quasiparticle) has a $-2e$ charge due to attractive interaction, and 
that the fundamental flux period of a material is equal to that of the
quasiparticle in the system ($2\pi/2e$).

\begin{figure}[htbp]
 \begin{center}
  \psfrag{a}{(a)}
  \psfrag{b}{(b)}
  \psfrag{p}{$\Phi$}
  \psfrag{I}{$I$}
  \includegraphics[scale=0.6]{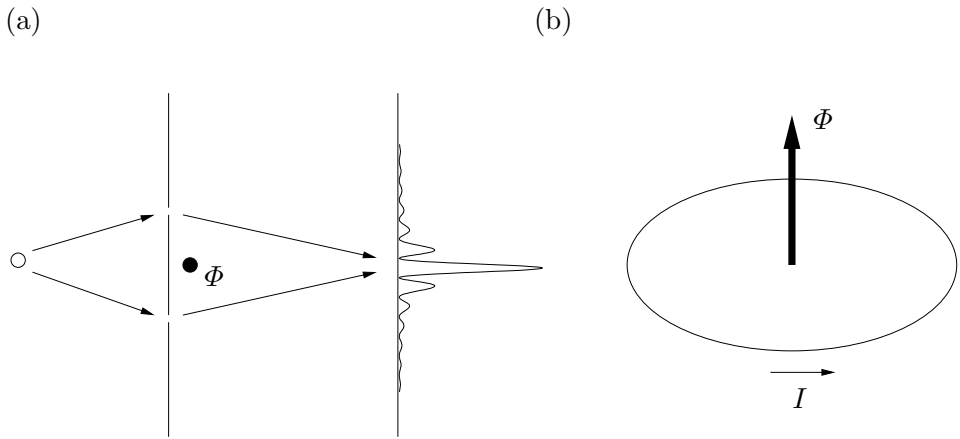}
 \end{center}
 \caption{Schematic diagram of (a) an Aharonov-Bohm experiment and (b)
 Aharonov-Bohm ring. 
 In (a), electrons are emitted from the source region, denoted by the
 empty circle, and produce the interference pattern.
 In (b), one-dimensional ring is pierced by a magnetic field and
 exhibits a persistent current.} 
 \label{fig:AB}
\end{figure}

In Fig.~\ref{fig:AB}, we depict schematic diagrams of typical
experimental settings for the Aharonov-Bohm effect. 
Consider an electron going over and under a very long impenetrable
cylinder, denoted as a black circle in Fig.~\ref{fig:AB}(a). 
In the cylinder there is a magnetic field parallel to the cylinder 
axis, taken as normal to the plane of the figure.
The probability of finding this electron in the interference region
depends on the magnetic field $\Phi$ and the interference pattern having 
period $\Phi_0$.
When an electron is placed in a ring pierced by a magnetic flux $\Phi$
as shown in Fig.~\ref{fig:AB}(b), a current $I$ defined by
differentiating the ground state energy with respect to the magnetic
flux, appears and is expected to have the flux periodicity $\Phi_0$.

Actual mesoscopic rings contain many electrons, and non vanishing
currents, known as persistent currents~\cite{BIL,Webb,Imry}, are also
expected.
Although the ground state in mesoscopic rings consists of many
electrons, because each electron has the flux periodicity $\Phi_0$ one 
might think the flux periodicity of the systems is the same as the
flux periodicity of the constituent electron. 
In this paper, we present an example in which the fundamental flux
period of the ground-state does not coincide with the flux periodicity
of the constituent particle. 
We will show that a fractional flux periodicity ($\Phi_0/2,\Phi_0/3,
\cdots$) can be realized in a certain limit of planar systems made from
a square lattice whose geometry is a torus~\footnote{It is noted that
fractional ($\Phi_0/3,\Phi_0/4,\cdots$ except $\Phi_0/2$) flux
periodicity presented in this paper holds in the long length and large
diameter limit of a torus.}.
To analyze the flux periodicity of the systems, we consider the ground
state energy and persistent currents.
These are considered suitable for examining the flux periodicity of the
ground state since they are regarded as the Aharonov-Bohm effect in
solid state systems~\cite{Imry}.
In the previous paper~\cite{SKS-lett}, we showed a typical example of a
fractional periodicity in the Aharonov-Bohm effect.
Here, we extend the theory in general cases and give detailed derivation
of the results.

This paper is organized as follows.
In Section~\ref{ground-state energy}, we obtain an explicit formula for
the ground state energy and plot it for different lattice structures,
showing flux periodicity numerically.
In Section~\ref{sec:persist}, we examine the persistent currents and
prove fractional flux periodicity in the long length and large diameter
limit of a torus noting that the characteristic features of the ground
state energy can be understood in terms of the twist-induced gauge
field.
In Section~\ref{sec:sum-dis}, we give a discussion and summary of our
results. 
In Appendix~\ref{app:pc}, we prove a formula used in the text and
estimate a correction to a fractional periodicity.
Throughout this paper we ignore electron spin and the effect of finite
temperature, and consider a half-filling system.

\section{Ground state energy}\label{ground-state energy}

\begin{figure}[htbp]
 \begin{center}
  \includegraphics[scale=0.3,angle=-90]{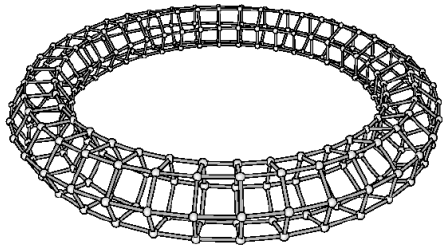}
 \end{center}
 \caption{Lattice structure of a twisted torus. 
 The lattice configuration is given by $N=6$, $Q = 40$ and $\delta N = 1$.}
 \label{fig:t-torus}
\end{figure}

\begin{figure}[htbp]
 \begin{center}
  \psfrag{a}{(a)}
  \psfrag{b}{(b)}
  \psfrag{C_h}{$C_h=NT_x$}
  \psfrag{T_w}{$T_w=\delta NT_x + Q T_y$}
  \psfrag{k}{$\delta N$}
  \psfrag{x}{$x$}
  \psfrag{y}{$y$}
  \psfrag{u}{u}
  \psfrag{d}{d}
  \psfrag{T}{$\delta N$}
  \includegraphics[scale=0.4]{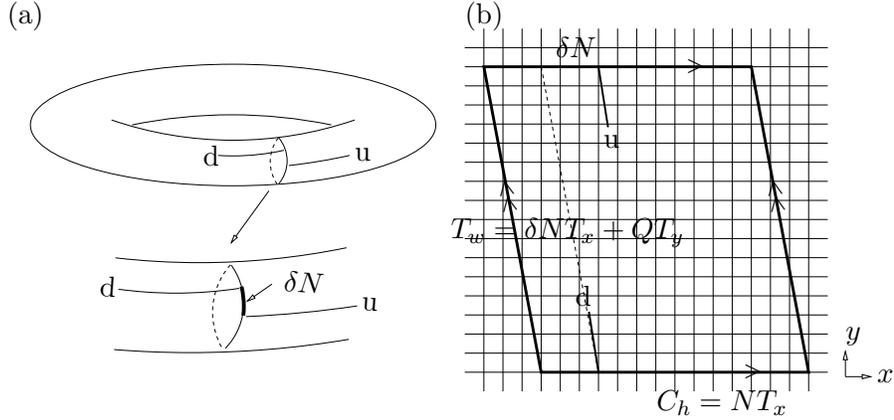}
 \end{center}
 \caption{Schematic diagram of (a) a twisted torus (upper) and the
 extension around a cross section (lower), and (b) its net diagram.
 It is convenient to consider a parallelogram sheet of square lattice as
 the net diagram of a twisted torus, where the chiral and translational
 vectors become the vectors at the sides of the parallelogram. 
 The lattice configuration is given by $N=14$, $Q = 16$ and $\delta N = -3$.
 } 
 \label{fig:torus}
\end{figure}

The lattice structure of a torus is specified by two vectors: the
chiral~\footnote{We borrow this terminology from the carbon nanotube
context~\cite{SDD}.}, given by $C_h = N T_x + M T_y$, and the
translational, given by $T_w = PT_x + QT_y$, where $N,M,P$ and $Q$ are
integers and $T_a$ ($a = x,y$) is the unit vector of the square lattice
in the direction of $x$ and $y$ where we define $x$ and $y$ in the two
dimensional map (see Fig.~\ref{fig:torus}(b)). 
Here, $C_h$ $(T_w)$ characterizes the lattice structure around (along)
the tube axis.  
We rewrite $T_a$ in terms of the chiral and translational vectors for a
later convenience as 
\begin{equation}
 \begin{pmatrix} T_x \cr T_y \end{pmatrix} = 
 \frac{1}{N_s} \begin{pmatrix} Q & -M \cr -P & N \end{pmatrix}
 \begin{pmatrix} C_h \cr T_w \end{pmatrix},
 \label{eq:index}
\end{equation}
where we define $N_s \equiv NQ-MP$, and $|N_s|$ corresponds to the
number of square lattices on the surface of the torus given by $C_h
\times T_w = N_s (T_x \times T_y)$.
In this paper, we take the $x$ axis in the direction of $C_h$ ($M = 0$),
and denote $P =\delta N$ to describe the {\it twist} around the tube
axis. 
Then, $C_h \cdot T_w = \delta N N a^2$ where $a(=|T_x|=|T_y|)$ is the
lattice constant.
We depict an example of twisted torus in Fig.~\ref{fig:t-torus}, in
which all sites are connected by a line consisting nearest-neighbor
bond.
Fig.~\ref{fig:torus}(a) shows an illustration of the twisted torus.
A torus can be unrolled to a parallelogram sheet as shown in
Fig.~\ref{fig:torus}(b).
In the figure, the two lines extending upward from `u' and downward from
`d' and having the same `$x$' at the junction are not joined for a
twisted torus, as shown in the inset of Fig.~\ref{fig:torus}(a).
There are $\delta N$ square lattices between the two lines.

We consider the quantum mechanical behavior of conducting electrons on a
square lattice using the nearest-neighbor tight-binding Hamiltonian: 
\begin{equation}
 {\cal H} = -t \sum_i \sum_{a=x,y} a_{i+a}^\dagger 
  e^{-ie A^{\rm ex} \cdot  T_a} a_i + h.c.,
\end{equation}
where $t$ is the hopping integral, $A^{\rm ex}$ is a constant external
gauge field, and $a_i$ and $a_i^\dagger$ are canonical
annihilation-creation operators of the electrons at site $i$ that
satisfy the anti-commutation relation $\{ a_i,a_j^\dagger \} =
\delta_{ij}$.
We diagonalize ${\cal H}$ in terms of the Bloch basis parametrized by
wave vector $k$ and obtain
${\cal H} = \sum_k E(k-eA^{\rm ex}) a_k^\dagger a_k$,
where $a_k \equiv \frac{1}{\sqrt{|N_s|}} \sum_i e^{ik \cdot r_i}a_i$ is
the annihilation operator of the Bloch orbital.
The energy eigenvalue of the Hamiltonian $E(k-eA^{\rm ex})$ is given by,
\begin{equation}
  E(k-eA^{\rm ex}) = 
  -2t \Re\left( \sum_{a=x,y} e^{i(k-eA^{\rm ex}) \cdot T_a} \right).
  \label{eq:eigenvalue1}
\end{equation}
The Bloch wave vector satisfies the periodic boundary condition for
$C_h$ and $T_w$ through which the geometrical (or topological)
information of a torus, such as the twist, is entered into the energy
eigenvalue. 
We decompose wave vector $k$ as $\mu_1 k_1 + \mu_2 k_2$, where $k_1$ and
$k_2$ are defined by
\begin{equation}
 C_h \cdot k_1 = 2\pi,\ C_h \cdot k_2 = 0,\
 T_w \cdot k_1 = 0 \ \text{and} \ T_w \cdot k_2 = 2\pi.
 \label{eq:constraint}
\end{equation}
Here, $k_1$ and $k_2$ can be rewritten in terms of the reciprocal
lattice vectors $K_x$ and $K_y$ as
\begin{equation}
 k_1 = \frac{1}{N_s} (QK_x -\delta N K_y), \ \ \ 
  k_2 = \frac{1}{N_s} (N K_y),
  \label{eq:fermi-point}
\end{equation}
where the reciprocal lattice vectors are defined by $K_a \cdot T_b =
2\pi \delta_{ab}$ with $a,b=x,y$.
Components $\mu_1$ and $\mu_2$ are quantum numbers (integers) and define
the Brillouin zone as
\begin{equation}
 \left[ - \frac{N}{2} \right] +1 \le \mu_1 \le \left[ \frac{N}{2} \right],\ \ \ 
  \left[ - \frac{Q}{2} \right] +1 \le \mu_2 \le \left[ \frac{Q}{2} \right],
\end{equation}
where $[n]$ indicates the maximum integer smaller than $n$.

We point out that the wave vector shifts according to the twist of a
torus.
For example, in the absence of twist ($\delta N = 0$), $k_1$ and $k_2$
are proportional to $K_x$ and $K_y$, respectively and perpendicular to
each other; otherwise, they are not. 
The shift of the wave vector plays an important role when we consider
the ground state configuration, which we discuss below.
Here we rewrite Eq.(\ref{eq:eigenvalue1}) by means of
Eqs.(\ref{eq:index}) (with $M=0$ and $P=\delta N$) and 
(\ref{eq:constraint}) as
\begin{equation}
 E(k-eA^{\rm ex}) = 
 -2t \ \left\{
      \cos \left( \frac{2\pi \mu_1}{N} \right) + 
      \cos \left( \frac{2\pi (\mu_2-\frac{\delta N}{N} \mu_1) 
	    - eA^{\rm ex} \cdot T_w }{Q} \right)
	  \right\},
 \label{eq:eigen-1}
\end{equation}
where we assume $A^{\rm ex} \cdot C_h =0$, and $A^{\rm ex} \cdot T_w$
corresponds to the Aharonov-Bohm flux penetrating the center of the
torus.
\begin{figure}[htbp]
 \begin{center}
  \psfrag{x}{$K_x$}
  \psfrag{y}{$K_y$}
  \psfrag{t1}{$T_x$}
  \psfrag{t2}{$T_y$}
  \psfrag{a}{$\frac{\pi}{a}$}
  \psfrag{k+}{$k^+$}
  \psfrag{k-}{$k^-$}
  \psfrag{s+}{$k_+$}
  \psfrag{s-}{$k_-$}
  \psfrag{s}{(a)}
  \psfrag{l}{(b)}
  \psfrag{X}{$\frac{K_x}{N}$}
  \psfrag{Y}{$\frac{K_y}{Q}$}
  \psfrag{m}{$\mu_1 > 0$}
  \psfrag{n}{$\mu_1 < 0$}
  \psfrag{A}{$-|\mu_1| \frac{\delta N}{N}$}
  \psfrag{B}{$+|\mu_1| \frac{\delta N}{N}$}
  \includegraphics[scale=0.5]{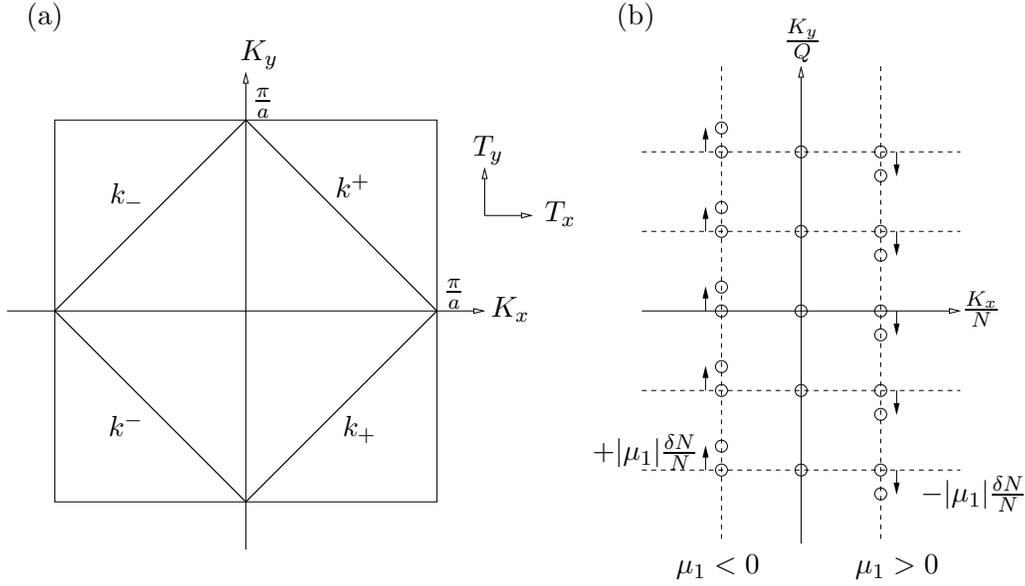}
 \end{center}
 \caption{Brillouin zone and Fermi lines in momentum space (a), and
 schematic diagram of the wave vector shifting due to the twist around
 the axis (b). There are two factors we should take into account to
 define the ground state.
 The first factor is the effect of the twist.
 Compared with the ground-state configuration in an untwisted torus,
 Eq.(\ref{eq:fermi-point}) indicates that the location of the energy
 eigenstates shifts depending on twist $\delta N$ and on the electron
 motion around the axis ($\mu_1$).
 This shift is illustrated in (b), where we denote the Bloch states as
 circles and depict several states of positive $\mu_1$, negative $\mu_1$
 and $\mu_1 = 0$. 
 In the absence of twist, all states are located at mesh line crossings.
 When twist is present, the states shift depending on their $\mu_1$ and
 $\delta N/N$.
 The second factor is the external gauge field. 
 The external gauge field can shift the wave vector, however, the shift is
 independent of electron motion contrary to the case where twist is
 present. }
 \label{fig:Fermi-line}
\end{figure}
The ground state is defined as the state for which all electronic states
below (above) the Fermi surface are occupied (empty). 
We consider a half-filling system and set the Fermi energy $E_{\rm
F}=0$.
We define the Fermi line in the two dimensional Brillouin zone, within
which all states are occupied, as the solution of $E(k) = 0$, forming a
square as shown in Fig.~\ref{fig:Fermi-line}(a). 
The side of the square (or Fermi line) is denoted by $k^\pm$ and
$k_\pm$.
Each energy eigenstate is labeled by $\mu_1$ and $\mu_2$ and thus we can
specify the occupied states by $\mu_1$ and $\mu_2$ explicitly.
In fact the region inside the Fermi line is defined as
\begin{align}
 & -\pi \le \frac{2\pi\mu_1}{N} + \frac{2\pi (\mu_2-\frac{\delta N}{N} \mu_1) 
 - eA^{\rm ex} \cdot T_w }{Q} \le \pi, \label{eq:r1} \\
 &  -\pi \le \frac{2\pi\mu_1}{N} - \frac{2\pi (\mu_2-\frac{\delta N}{N} \mu_1) 
 - eA^{\rm ex} \cdot T_w }{Q} \le \pi, \label{eq:r2}
\end{align}
where we note that the Fermi lines satisfy $k^{\pm} \cdot (T_x + T_y) =
\pm \pi$ and $k_\pm \cdot (T_x-T_y) = \pm \pi$, and Eqs.(\ref{eq:r1})
and (\ref{eq:r2}) can be rewritten as
\begin{align}
 & k^- \cdot (T_x + T_y) \le (k-eA^{\rm ex}) \cdot (T_x + T_y)  \le k^+ \cdot
 (T_x + T_y), \\
 & k_- \cdot (T_x -T_y) \le (k-eA^{\rm ex})\cdot (T_x -T_y) \le k_+ \cdot
 (T_x-T_y). 
\end{align}
For a fixed value of $\mu_1$, we obtain the following region of $\mu_2$,
in which all states are occupied by electrons: 
\begin{equation}
 \left[ -Q \left( \frac{1}{2} - \frac{\mu_1}{N} \right) + \frac{\delta N}{N} \mu_1 + N_\Phi \right] + 1
  \le \mu_2 \le 
  \left[ Q \left( \frac{1}{2} - \frac{\mu_1}{N} \right) + \frac{\delta N}{N} \mu_1 + N_\Phi \right], \label{eq:bri-1}
\end{equation}
for $\mu_1 \ge 0$, and 
\begin{equation}
 \left[ -Q \left( \frac{1}{2} + \frac{\mu_1}{N} \right) + \frac{\delta N}{N} \mu_1 + N_\Phi \right] + 1
 \le \mu_2 \le 
 \left[ Q \left( \frac{1}{2} + \frac{\mu_1}{N} \right) + \frac{\delta
 N}{N} \mu_1+ N_\Phi \right], \label{eq:bri-2}
\end{equation}
for $\mu_1 \le 0$, respectively.
Here, we set $A^{\rm ex} \cdot T_w = \Phi$ and define the number of
Aharonov-Bohm flux $N_\Phi \equiv \Phi/\Phi_0$.
The ground-state energy is therefore
\begin{equation}
 E_0 (\Phi) = \sum_{\mu_1=\left[-\frac{N}{2}\right]+1}^{\left[\frac{N}{2}\right]} 
  \sum_{\mu_2=\left[-Q\left( \frac{1}{2} - \frac{|\mu_1|}{N} \right) + \frac{\delta N}{N} \mu_1 + N_\Phi \right]+1}^{\left[Q\left( \frac{1}{2} - \frac{|\mu_1|}{N} \right) + \frac{\delta N}{N} \mu_1 + N_\Phi \right]}
  E(k -eA^{\rm ex}).
  \label{eq:vac-ene}
\end{equation}
This is the explicit mathematical expression for the ground state energy
of a twisted torus. 

\begin{figure}[htbp]
 \begin{center}
  \psfrag{x}{$N_\Phi$}
  \psfrag{y}{Ground state energy $E_0(\Phi)/4t$}
  \includegraphics[scale=0.8]{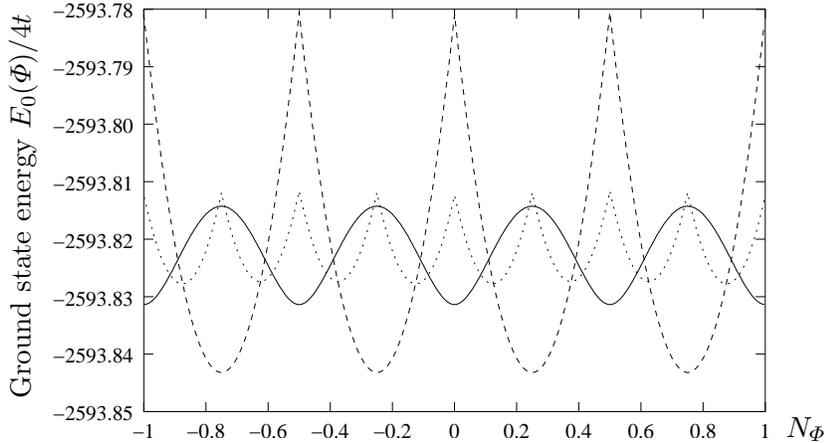}
 \end{center}
 \caption{Ground state energy as a function of flux.
 We fix $N=80$ and $Q=160$, then plot for $\delta N = 1,20,40$ to
 examine the dependence of the ground-state energy on twist.
 The observed flux periodicity seems to be $\Phi_0/2$ for $\delta N = 1$
 (solid line) and $\delta N = 40$ (dashed line) and $\Phi_0/4$ for 
 $\delta N = 20$ (dotted line).
 For the untwisted torus ($N=40$, $Q=160$ and $\delta N =0$), refer to   
 Fig.~\ref{fig:g-ene-Q160}.} 
 \label{fig:g-ene-dN}
\end{figure}
Here, we plot $E_0(\Phi)/4t$ for different $N$, $\delta N$ and $Q$ as a
function of $N_\Phi$, and observe the characteristics of the results.
Fig.~\ref{fig:g-ene-dN} shows the dependence of the ground state energy
on $\delta N$, where we fix $N = 80$ and $Q=160$ and vary the twist as
$\delta N = 1,20,40$. 
The flux periodicity seems to be $\Phi_0/2$ for $\delta N = 1$ (solid
line) and $\delta N = 40$ (dashed line), and $\Phi_0/4$ for $\delta N =
20$ (dotted line).
It appears that flux periodicity of the dashed and dotted lines
corresponds to the lattice structure of $\delta N/N$. 
We note that $\Phi_0/2$ periodicity of solid line ($\delta N = 1$: odd
number) is accurate in our numerical data (10-figure numbers) and that
of dashed line ($\delta N = 40$: even number) is approximate one.
In addition to observing flux periodicity, we see that the solid line is
smooth (no cusp) at $N_\Phi =0$ for example, when compared with the
other two lines.

\begin{figure}[htb]
 \parbox{\halftext}{
 \psfrag{x}{$N_\Phi$}
 \psfrag{y}{$(E_0(\Phi)-E_0(0))/4t$}
 \subfigure[Ground state energy for $N=40$, $Q=160$ and $\delta N =20$. 
 We plot $(E_0(\Phi)-E_0(0))/4t$ where $E_0(0)/4t = -1296.892$.]
 {\includegraphics[scale=0.5]{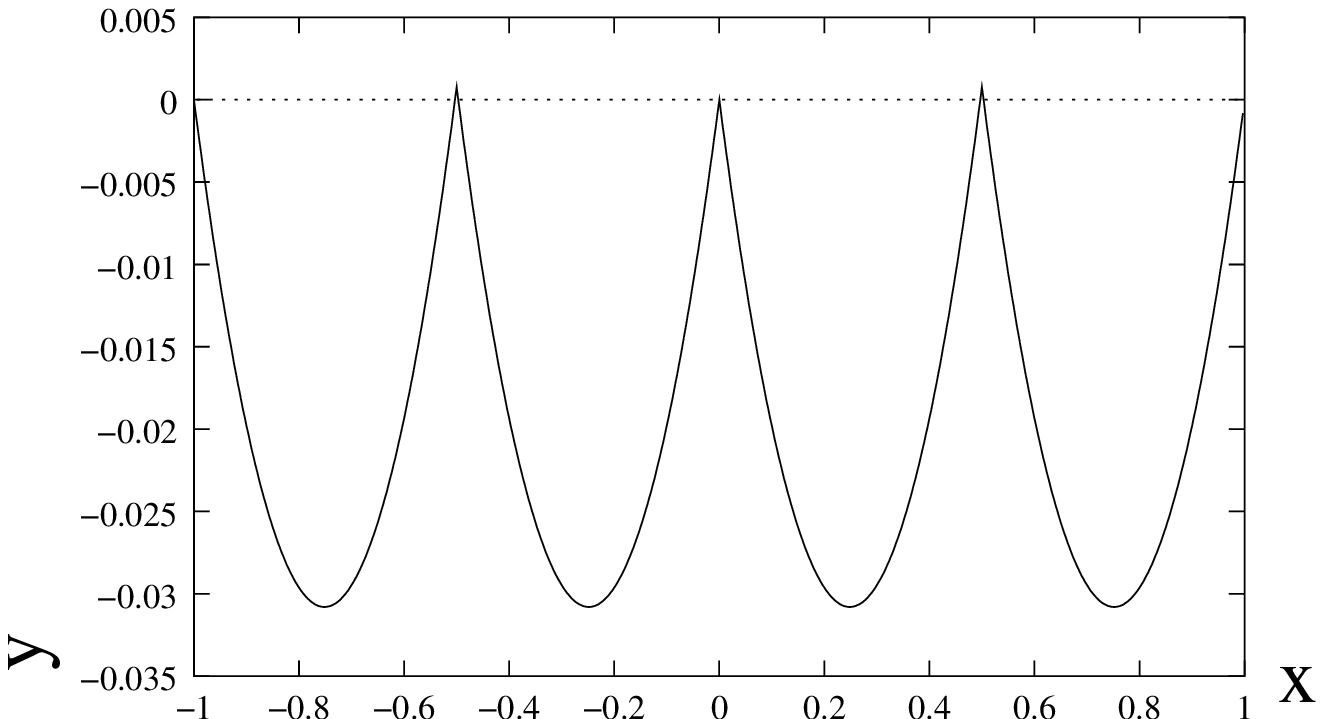} \label{fig:g-ene-N40}}}
 \hfill
 \parbox{\halftext}{
 \psfrag{x}{$N_\Phi$}
 \psfrag{y}{$(E_0(\Phi)-E_0(0))/4t$}
 \subfigure[Ground state energy for $N=80$, $Q=160$ and $\delta N =20$.
 We plot $(E_0(\Phi)-E_0(0))/4t$ where $E_0(0)/4t = -2593.812$.]{\includegraphics[scale=0.5]{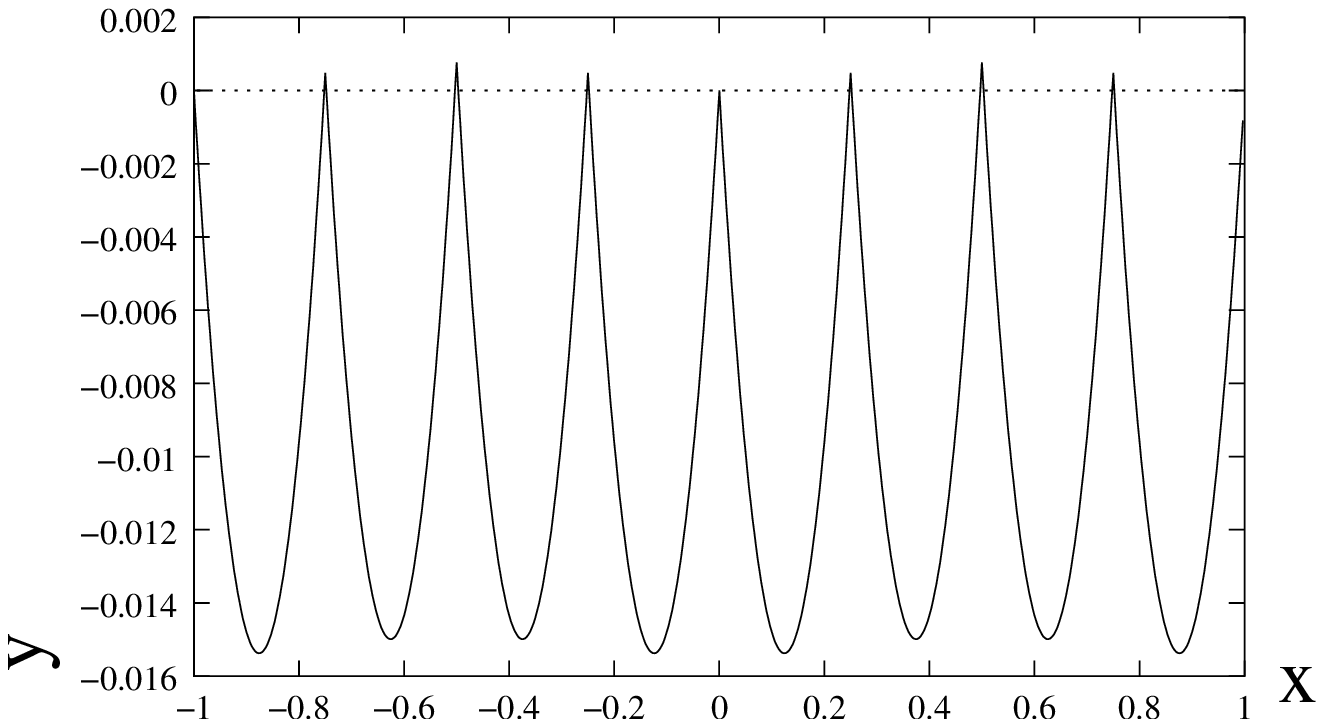}\label{fig:g-ene-N80}}}
 \parbox{\halftext}{
 \psfrag{x}{$N_\Phi$}
 \psfrag{y}{$(E_0(\Phi)-E_0(0))/4t$}
 \subfigure[Ground state energy for $N=81$, $Q=160$ and $\delta N =20$.
 We plot $(E_0(\Phi)-E_0(0))/4t$ where $E_0(0)/4t = -2626.2443$.]{\includegraphics[scale=0.5]{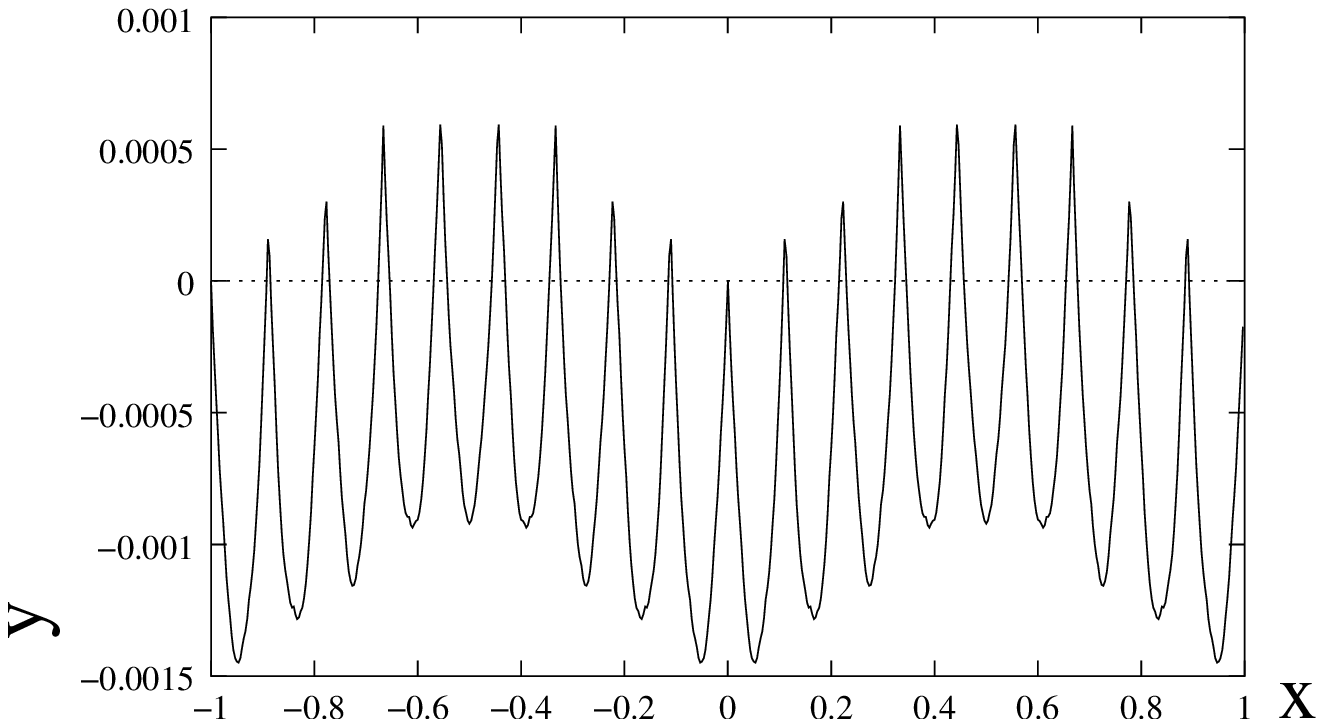}\label{fig:g-ene-N81}}}
 \hfill
 \parbox{\halftext}{
 \psfrag{x}{$N_\Phi$}
 \psfrag{y}{$(E_0(\Phi)-E_0(0))/4t$}
 \subfigure[Ground state energy for $N=160$, $Q=160$ and $\delta N =20$.
 We plot $(E_0(\Phi)-E_0(0))/4t$ where $E_0(0)/4t = -5187.640$.]
 {\includegraphics[scale=0.5]{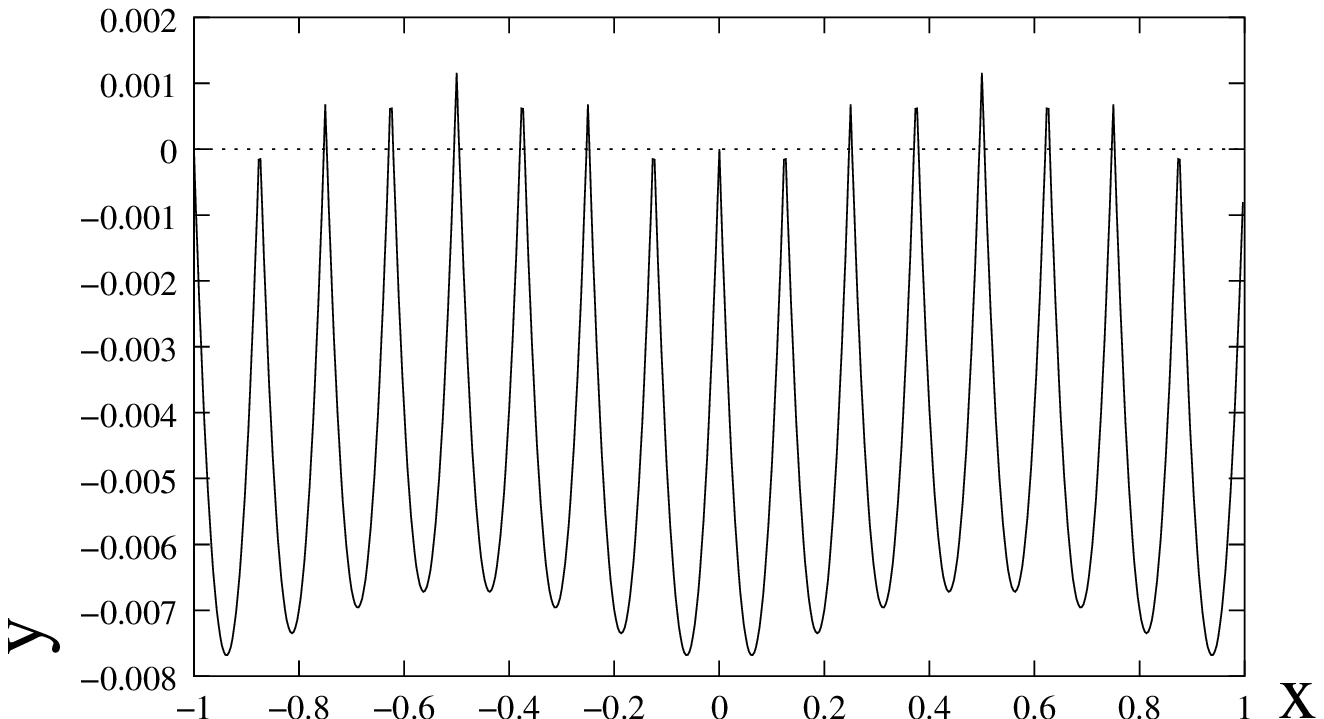}\label{fig:g-ene-N160}}}
 \caption{Dependence of the ground state energy on the diameter of the
 tube $N$.} 
\end{figure}

Next we examine the dependence of the ground state energy on the
diameter of the tube $N$.
Figs.~\ref{fig:g-ene-N40},~\ref{fig:g-ene-N80}~\footnote{This figure
is a refined version of the dotted line in
Fig.~\ref{fig:g-ene-dN}.},~\ref{fig:g-ene-N81} and 
\ref{fig:g-ene-N160} show the ground state energy with fixed $Q =160$
and $\delta N = 20$ while varying $N=40,80,81,160$. 
Observing the flux periodicity of the samples, they appear to be
periodic in Figs.~\ref{fig:g-ene-N40} and ~\ref{fig:g-ene-N80} with a
period of $\Phi_0/2$ in Fig.~\ref{fig:g-ene-N40} and $\Phi_0/4$ in
Fig.~\ref{fig:g-ene-N80}.
They are not periodic, however, due to a small modulation in the ground
state energy.
This is manifest in Figs.~\ref{fig:g-ene-N81} and ~\ref{fig:g-ene-N160}. 
This modulation can be thought of as a finite size correction which will
be discussed in Section~\ref{sec:persist}.

\begin{figure}[htb]
 \parbox{\halftext}{
 \psfrag{x}{$N_\Phi$}
 \psfrag{y}{$(E_0(\Phi)-E_0(0))/4t$}
 \subfigure[Ground state energy for $N=40$, $Q=160$ and $\delta N = 0$.
 We plot $(E_0(\Phi)-E_0(0))/4t$ where $E_0(0)/4t = -1296.83$.]
 {\includegraphics[scale=0.5]{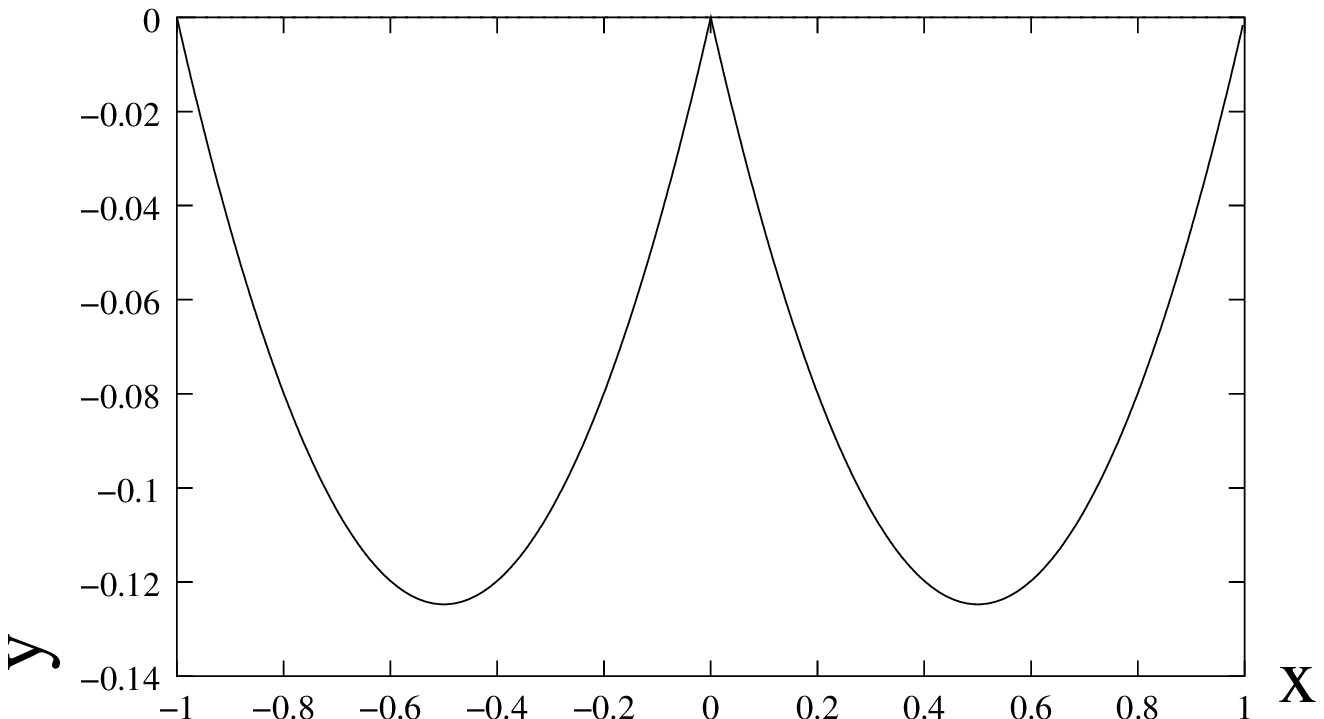}\label{fig:g-ene-Q160}}}
 \hfill
 \parbox{\halftext}{
 \psfrag{x}{$N_\Phi$}
 \psfrag{y}{$(E_0(\Phi)-E_0(0))/4t$}
 \subfigure[Ground state energy for $N=40$, $Q=170$ and $\delta N = 0$.
 We plot $(E_0(\Phi)-E_0(0))/4t$ where $E_0(0)/4t = -1377.965$.]
 {\includegraphics[scale=0.5]{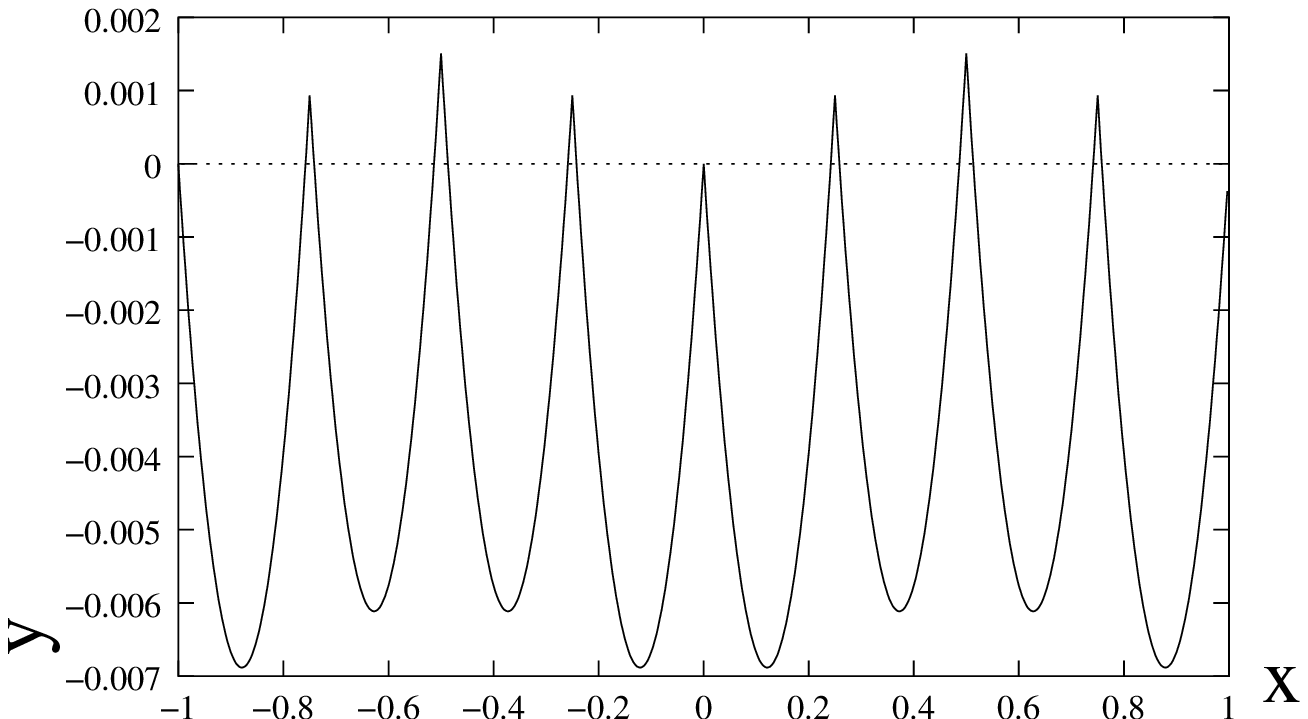}\label{fig:g-ene-Q170}}}
 \parbox{\halftext}{
 \psfrag{x}{$N_\Phi$}
 \psfrag{y}{$(E_0(\Phi)-E_0(0))/4t$}
 \subfigure[Ground state energy for $N=40$, $Q=180$ and $\delta N = 0$.
 We plot $(E_0(\Phi)-E_0(0))/4t$ where $E_0(0)/4t = -1459.008$.]
 {\includegraphics[scale=0.5]{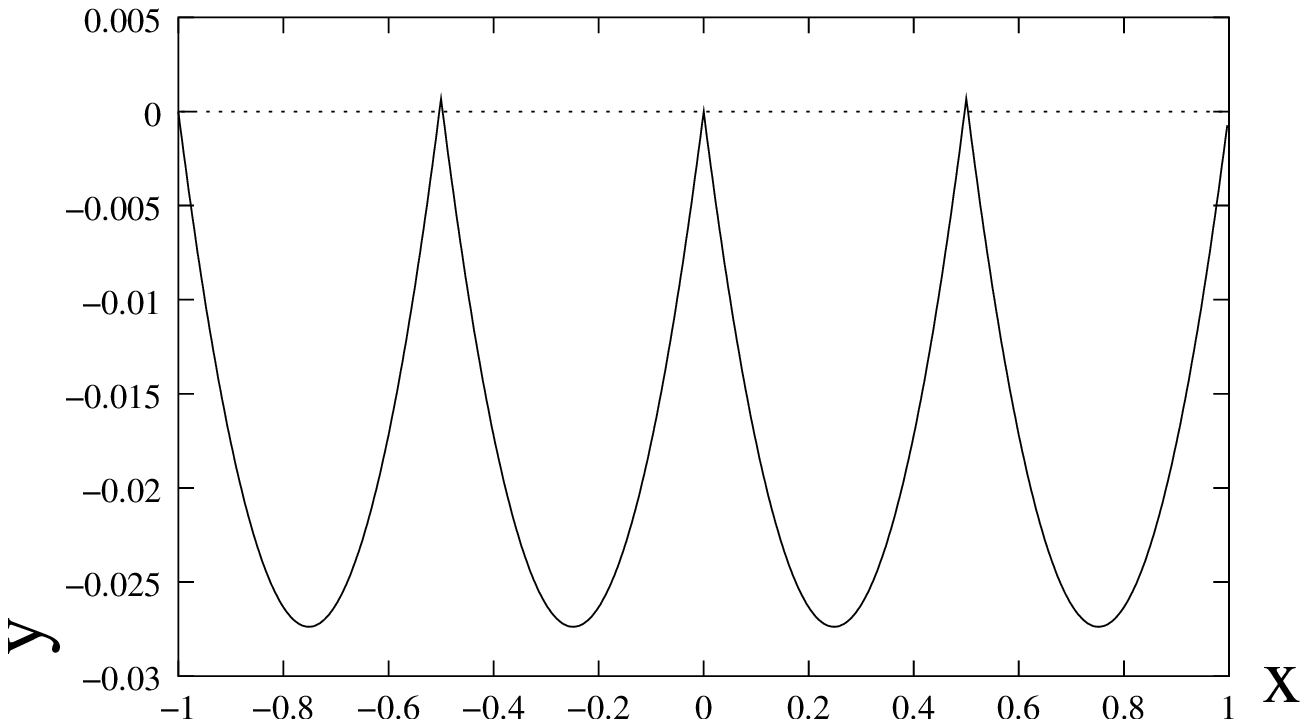}\label{fig:g-ene-Q180}}}
 \hfill
 \parbox{\halftext}{
 \psfrag{x}{$N_\Phi$}
 \psfrag{y}{$(E_0(\Phi)-E_0(0))/4t$}
 \subfigure[Ground state energy for $N=40$, $Q=181$ and $\delta N = 0$.
 We plot $(E_0(\Phi)-E_0(0))/4t$ where $E_0(0)/4t = -1467.135$.]
 {\includegraphics[scale=0.5]{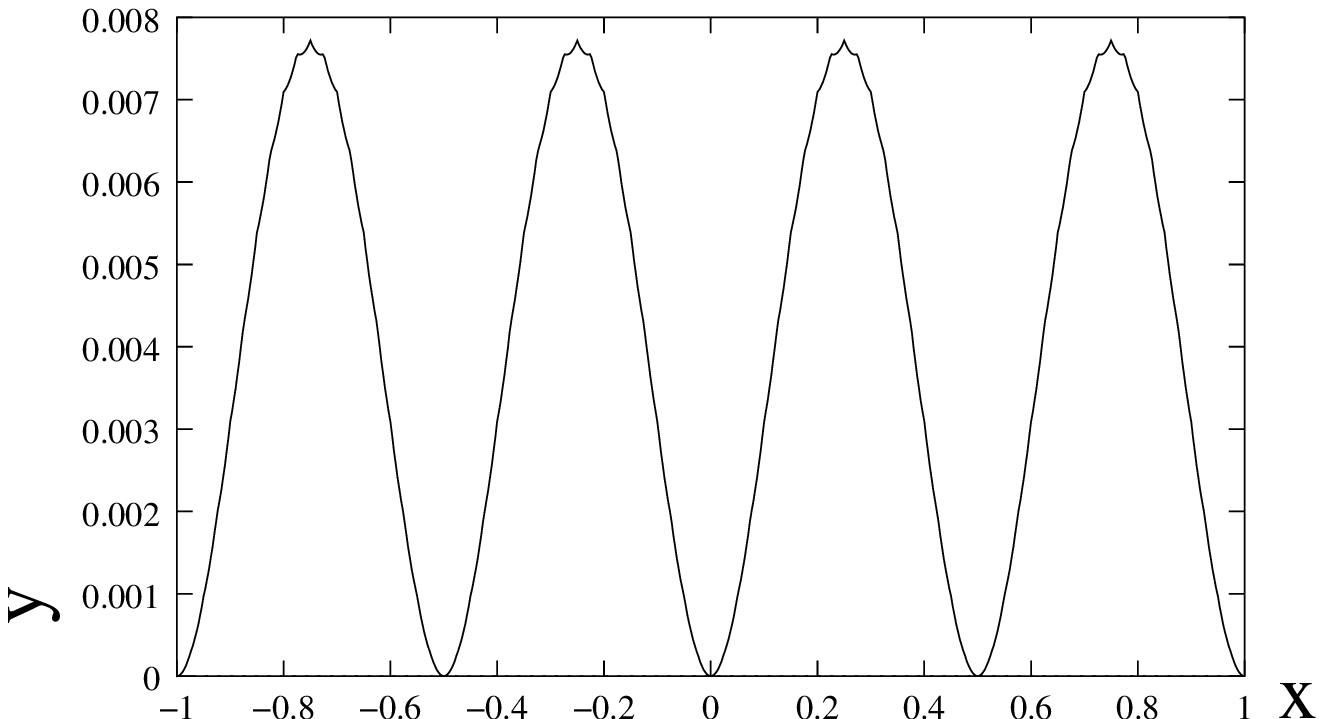}\label{fig:g-ene-Q181}}}
 \caption{Dependence of the ground state energy on the diameter of the
 torus $Q$.}
\end{figure}

Finally, we observe the dependence of the ground state energy on $Q$
shown in
Figs.~\ref{fig:g-ene-Q160},~\ref{fig:g-ene-Q170},~\ref{fig:g-ene-Q180} 
and~\ref{fig:g-ene-Q181}.
Here, we note that all previous examples except Fig.~\ref{fig:g-ene-N81}
have had the common feature of $Q$ as a multiple of $N$.
Fig.~\ref{fig:g-ene-Q160} shows the standard flux periodicity $\Phi_0$.
In Fig.~\ref{fig:g-ene-Q180}, however, we have an approximate
periodicity of $\Phi_0/2$ even in the absence of twist.
We note that $\Phi_0/2$ periodicity in Fig.~\ref{fig:g-ene-Q181}
($Q=181$: odd number) is accurate in our numerical data.

It is important to mention that flux periodicity and the amplitude of
the oscillation of $E_0(\Phi)$ depends much on $N$, $Q$ and $\delta N$.
We will discuss the dependence of the flux periodicity and amplitude on
them in the next section.

\section{Persistent currents and interpretation of the numerical
 results}\label{sec:persist}

In the previous section, we presented a formula for the ground state
energy. 
By plotting the energy for different values of $N,\delta N$ and $Q$, we
found that some tori exhibit an approximate fractional flux periodicity.
In this section, we will present an analytical results that demonstrate
and prove fractional flux periodicity in the long length ($Q \gg 1$)
and large diameter ($N \gg 1$) limit for a twisted torus in terms of the
persistent currents. 
By so doing, we clarify the relationship between the lattice structure
and flux periodicity.

Persistent current is defined by differentiating the ground state energy
with respect to the magnetic flux $\Phi$~\cite{Imry}:
\begin{equation}
 I_{\rm pc}(\Phi) = - \frac{\partial E_0(\Phi)}{\partial \Phi}.
\end{equation}
Because $E_0(\Phi)$ consists of many energy dispersion relations as
specified by $\mu_1$ (hereafter referred to as $\mu_1$-th energy band),
we can write 
\begin{equation}
 E_0(\Phi) =
  \sum_{\mu_1=\left[-\frac{N}{2}\right]+1}^{\left[\frac{N}{2}\right]}  
  E(\mu_1,\Phi)
\end{equation}
where $E(\mu_1,\Phi)$ is the ground state energy of $\mu_1$-th energy
band and is the sum of the energy at $\mu_2$-th $k$ points which satisfy
Eqs.(\ref{eq:bri-1}) and (\ref{eq:bri-2}):
\begin{equation}
 E(\mu_1,\Phi) \equiv 
  \sum_{\mu_2=\left[-Q\left( \frac{1}{2} - 
		       \frac{|\mu_1|}{N} \right) + 
  \frac{\delta N}{N} \mu_1 + N_\Phi \right]+1}^{\left[Q\left( \frac{1}{2}
						 - \frac{|\mu_1|}{N} \right) 
  + \frac{\delta N}{N} \mu_1 + N_\Phi \right]}
  E(\mu_1 k_1 + \mu_2 k_2 -eA^{\rm ex}).
  \label{eq:vac-ene-mu_1}
\end{equation}
Hence, the persistent current can also be decomposed into the current of
each energy band as
\begin{equation}
 I_{\rm pc}(\Phi) = 
  \sum_{\mu_1= \left[ -\frac{N}{2} \right]+1}^{\left[ \frac{N}{2} \right]}
  I(\mu_1, \Phi),\  \text{ where $I(\mu_1, \Phi) \equiv
  - \frac{\partial E(\mu_1,\Phi)}{\partial \Phi}$}.
  \label{eq:pc-mu_1}
\end{equation}
To calculate the persistent currents in long systems with $|T_w| \gg
a$ (or $Q \gg 1$), it is not necessary to sum over the energy
eigenvalue of the valence electrons but just calculate the Fermi
velocity of the energy band.
Because, when we vary the magnetic field, the change of the ground state
energy is almost determined by the change of the energy eigenvalue of
electrons near the Fermi level and that of the energy eigenvalues deep
in energy bands cancel to one another.
Using Fermi velocity, the amplitude of $I(\mu_1, \Phi)$ (the persistent
current for the $\mu_1$-th energy band) is well approximated
by~\cite{Imry} 
\begin{equation}
 I(\mu_1)=\frac{e v_F(\mu_1)}{|T_w|},
  \label{eq:pc-amp}
\end{equation}
where $|T_w|$ is the system length and $v_F(\mu_1)$ denotes the Fermi
velocity of the $\mu_1$-th energy band.
The proof of this approximation is given in Appendix~\ref{app:pc}.
By fixing $\mu_1$ and expanding Eq.(\ref{eq:eigen-1}) on $\mu_2$ around
the Fermi level, we obtain the energy dispersion relation of the
$\mu_1$-th energy band as
\begin{equation}
 {\cal H}_{\mu_1} = v_F(\mu_1) p_2 + {\cal O}(p_2^2).
  \label{eq:dispersion}
\end{equation}
where $p_2 (\equiv \mu_2 k_2)$ denotes the momentum along the axis. 
Here, we neglect an external gauge field.
Notice that Eq.(\ref{eq:dispersion}) represents the case for $\mu_2 >
0$, and for $\mu_2 < 0$, the energy dispersion relation is given by
$-{\cal H}_{\mu_1}$ up to ${\cal O}(p_2)$.
The coefficient of $p_2$ is the Fermi velocity and is obtained as
\begin{equation}
 v_F(\mu_1) = 2ta \left|\sin \left(\frac{2\pi \mu_1}{N} \right)\right|.
  \label{eq:fermi-vel-eff-mass}
\end{equation}

First we consider persistent currents in an untwisted torus ($\delta N
=0$).
In this case, all energy bands produce the same function for the
persistent current, which is a saw-tooth curve as a function of $\Phi$
with the same zero points (the flux for which the amplitude of the
current vanishes) but with different amplitudes.
In fact, the amplitude of the total current is given by a summation of
all amplitudes:
\begin{align}
 I_{\rm tot} =  \sum_{\mu_1=\left[ -\frac{N}{2} \right]+1}^{\left[ \frac{N}{2} \right]} 
  I(\mu_1) = 
  \begin{cases}
   \displaystyle \frac{2 eta}{|T_w|} \cot \frac{\pi}{N} & \text{ for $N = {\rm even}$,} \\
   \displaystyle \frac{eta}{|T_w|} \cot \frac{\pi}{2N}
   & \text{ for $N = {\rm odd}$.} 
  \end{cases}
 \label{eq:pc-untwist}
\end{align}
Here, we implicitly assume $Q$ is a multiple of $N$ (see
Appendix~\ref{app:pc}). 
The other cases will be mentioned in the end of this Section. 
Persistent current $I_{\rm pc}$ in the torus is given by $I_{\rm tot}$
multiplied by $-\phi/\pi$ ($\phi \equiv 2\pi
(\Phi/\Phi_0)$)~\footnote{We may multiply $I_{\rm tot}$ by $\phi/\pi$ as
well. The sign ($\pm \phi$) depends on the choice of positive direction 
along the torus axis and does not concern flux periodicity.}
The linear relation between $\Phi$ and the persistent currents has a
periodicity of $\Phi_0$, and we then have a saw-tooth current: 
\begin{align}
 I_{\rm pc}^{\delta N = 0}(\phi) = 
 - I_{\rm tot} \frac{2}{\pi} 
 \sum_{n = 1}^\infty (-1)^{n+1} \frac{\sin(n\phi)}{n}.
 \label{eq:pc-untwist}
\end{align}
It is noted that $\Phi_0$ fundamental periodicity obtained here is not
an approximate periodicity but is exact, and also that persistent
currents in a torus contain always $\Phi_0$ periodicity. 
However, the $\Phi_0$ periodicity is not always a fundamental
periodicity as is shown below.

We comment on the functional form of the persistent currents.
The saw-tooth persistent current is a characteristic of the
non-interacting theories at zero temperature.
Each saw-tooth curve loses its sharpness due to disorder, or at finite
temperature~\cite{Buttiker}.
We note here, and in all subsequent discussions, that we are implicitly
assuming the amplitude of the persistent current vanishes at $\phi =0$.
Generally, the zero amplitude points depend on the total number of
electrons in the system and may be $\phi= 2\pi j$ or $\phi=2\pi j+\pi$
where $j$ is an integer (Appendix~\ref{app:pc}). 
As for the flux periodicity, the zero amplitude position does not affect
the results. 
We note also that the persistent current corresponding to the ground
state energy shown in Fig.~\ref{fig:g-ene-Q160} can be approximated by
Eq.(\ref{eq:pc-untwist}) with the replacement $\phi \to \phi-\pi$.

Next, we consider a twisted torus ($\delta N \ne 0$).
To understand the effect of twist on the persistent currents, we
introduce {\it twist-induced} gauge field $A^{\rm twist}$.
In terms of $A^{\rm twist}$, the second term of the right hand side of 
Eq.(\ref{eq:eigen-1}) can be rewritten as  
\begin{equation}
 -2t \cos \left(\frac{2\pi \mu_2  
       - (eA^{\rm ex}+\mu_1 A^{\rm twist}) \cdot T_w }{Q} \right), 
\end{equation}
where we define $A^{\rm twist} \cdot C_h \equiv 0$ and $A^{\rm twist}
\cdot T_w \equiv 2\pi \delta N/N$.  
Like $A^{\rm ex}$, $A^{\rm twist}$ can shift the wave vector.
However, the shift depends on the twist ($\delta N/N$) and the wave
vector around the tube axis ($\mu_1 k_1$).
Because of this, coupling between $A^{\rm twist}$ and the conducting
electron preserves the time-reversal symmetry of the whole system. 
This is contrasted with the time-reversal symmetry broken by $A^{\rm
ex}$. 
The twist behaves as an extra gauge field and shifts the wave vector
along the axis direction, so that the zero points of the persistent
current (or a saw-tooth curve) also shift~\cite{Sasaki}.
As a result, in order to calculate the total current, we must sum the
saw-tooth curves that have different zero points and different
amplitudes, all of which depend on $\mu_1$.
Then we have
\begin{align}
 I_{\rm pc}^{\delta N}(\phi) = 
 - \frac{2eta}{|T_w|} I^{\delta N}_N(\phi), 
 \label{eq:pc-twist}
\end{align}
where $I^{\delta N}_N(\phi)$ is expressed as
\begin{align}
 I^{\delta N}_N(\phi) 
  &=
  \frac{2}{\pi} \sum_{\mu_1=\left[ -\frac{N}{2} \right]+1}^{\left[ \frac{N}{2}\right]} 
  \sum_{n = 1}^{\infty} (-1)^{n+1} 
  \frac{\sin \left(n(\phi-2\pi \mu_1 \frac{\delta N}{N}) \right)}{n}
  \left|\sin \left(\frac{2\pi \mu_1}{N}\right)\right| \nn \\
 &=
  \frac{1}{\pi} \sum_{n = 1}^{\infty} (-1)^{n+1}\frac{\sin(n\phi)}{n}
   \left( 1+\cos(n\pi \delta N) \right)
   \left( \frac{2\sin \left(\frac{2\pi}{N}\right)}{\cos \left(\frac{2\pi n \delta N}{N}
		       \right) -\cos \left(\frac{2\pi}{N}\right) } \right),
  \label{eq:pc-func-t}
\end{align}
where we assume $N$ is an even number.
When $N$ is an odd number, we should add the following extra term to the
right-hand side of Eq.(\ref{eq:pc-func-t}):
\begin{align}
 \frac{1}{\pi} \sum_{n = 1}^{\infty} (-1)^{n+1}\frac{\sin(n\phi)}{n}
 \cos(n\pi \delta N) \left( \frac{2\sin 
 \left(\frac{\pi}{N}\right)}{\cos \left(\frac{\pi n \delta N}{N}
 \right) + \cos \left(\frac{\pi}{N}\right) } \right).
 \label{eq:odd-N-finite}
\end{align}
Depending on an even or odd number of $\delta N$, term $1 + \cos(\pi n
\delta N)$ in Eq.(\ref{eq:pc-func-t}) vanishes if $n\delta N$ is an odd
number or is equal to 2 for other cases.
Then Eq.(\ref{eq:pc-func-t}) reduces to 
\begin{equation}
 I_{N}^{\delta N}(\phi) = 
  \begin{cases}
   \displaystyle -\frac{1}{\pi} \sum_{n = 1}^{\infty} \frac{\sin(2n\phi)}{n} C^{o}_n
   & \text{ for $\delta N = {\rm odd}$}, \\ 
   \displaystyle \frac{2}{\pi} \sum_{n = 1}^{\infty} \frac{(-1)^{n+1}\sin(n\phi)}{n}
   C^{e}_n & \text{ for $\delta N = {\rm even}$},
  \end{cases} \label{eq:pc-even}
\end{equation}
where $C^{o}_n$ and $C^{e}_n$ are given
\begin{equation}
 C^{o}_n = 
  \frac{2\sin \left(\frac{2\pi}{N}\right)}{\cos \left(\frac{4\pi n \delta N}{N}\right) 
  -\cos \left(\frac{2\pi}{N}\right) }, \ \
  C^{e}_n =
  \frac{2\sin \left(\frac{2\pi}{N}\right)}{\cos \left(\frac{2\pi n \delta N}{N}\right)
  -\cos \left(\frac{2\pi}{N}\right) }.
  \label{eq:Coe}
\end{equation}
The difference in periodicity for $\phi$ is clear because $\sin (2n
\phi)$ has a period one half the fundamental unit of magnetic flux, or
$\Phi_0/2$.
We note that the one-half flux periodicity for an odd number of $\delta
N$ becomes exact if $N$ is an even number, which means that the
ground-state energy Eq.(\ref{eq:vac-ene}) has one-half flux periodicity
($Q$ is not necessary to be a large number).  
The solid line ($\delta N =1$) in Fig.~\ref{fig:g-ene-dN} is an example
of $\delta N =\text{odd}$ showing an exact one-half flux periodicity.
This conclusion is valid for any even number of $N$.
For an odd number of $N$, exact $\Phi_0/2$ periodicity is absent when
$N$ is a finite number because of Eq.(\ref{eq:odd-N-finite}).
This accurate one-half periodicity is due to the fact that the
persistent current is an odd function (see Appendix~\ref{app:pc}) in the
absence of twist, which provides the term $1 + \cos(\pi n \delta N)$ in
the persistent current even for a finite $Q$ value.

A {\it fundamental} flux periodicity for a current demonstrates a more
interesting behavior for a specific combination of $N$ and $\delta N$.
For a while, let us fix $N$ an even number.
We consider a particular structure for $\delta N/N = 1/3$ where $\delta
N$ is an even number.
In this case, the fundamental period of $C^e_n$ is 3, i.e., $C^e_{n+3} =
C^e_n$, which shows there may exist $\Phi_0/3$ fundamental period in the
currents. 
In fact, when $N \gg 1$, we have $C^e_1 = C^e_2 \approx -8\pi/3N$ and
$C^e_3 \approx 2N/\pi$.   
We then obtain 
\begin{align}
 I_{N}^{\delta N}(\Phi)
 &=
 \frac{2}{\pi}  \sum_{n = 0}^{\infty} \sum_{m=1}^3
 (-1)^{3n+m+1}\frac{\sin((3 n+m)\phi)}{3 n+m} C^e_m \nn \\
 &=
 \frac{2}{3\pi} \sum_{n = 1}^{\infty} 
 (-1)^{n+1} \frac{\sin(3 n\phi)}{n} C^e_3 + {\cal O}(1/N) \nn \\
 &\approx
 \frac{2N}{\pi^2} \phi \ \  \text{for $N \gg 1$},
 \label{eq:1/3}
\end{align}
where in the last line of Eq.(\ref{eq:1/3}) we assume $-\pi/3 < \phi <
\pi/3$.
The fundamental flux period of Eq.(\ref{eq:1/3}) becomes $\Phi_0/3$.
It should be noted that this argument can be applied to other twisted
structures.
For instance, when $\delta N/N  = 1/Z$ ($1/Z=1/4,1/5,\cdots$),
we obtain another fractional period of $\Phi_0/Z$.
We note that the total persistent currents are still saw-tooth curves as
expected from the non-interacting theories.
The ground state energy corresponding to $Z=2$ and $Z=4$ is plotted in
Fig.~\ref{fig:g-ene-dN} as dashed and dotted lines, where there remains
a finite energy modulation which prevents an exact fractional
periodicity. 
It is noted that, in the large diameter limit ($N \gg 1$), the same
behavior holds for an odd number of $N$.

\begin{figure}[htbp]
 \begin{center}
  \psfrag{a}{$-\pi$}
  \psfrag{b}{$-\pi/2$}
  \psfrag{c}{$0$}
  \psfrag{d}{$\pi/2$}
  \psfrag{e}{$\pi$}
  \psfrag{x}{$\phi$}
  \psfrag{y}{$I^{\delta N}_{\rm pc}(\Phi)/\frac{|C_h|}{|T_w|} \frac{-2et}{3\pi^2\delta N^2}
$}
  \psfrag{E}{even}
  \psfrag{O}{odd}
  \includegraphics[scale=0.8]{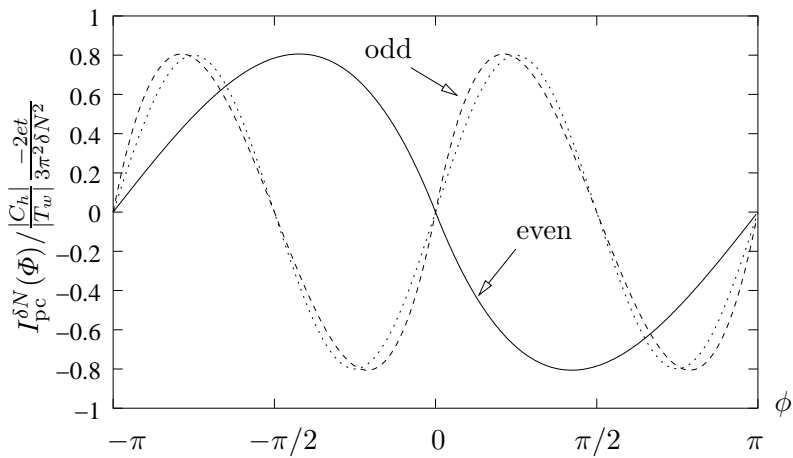}
 \end{center}
 \caption{Persistent currents as $N \to \infty$ (Eq.\ref{eq:limit}) for
 an even (solid line) and an odd (dashed line) number of $\delta N$. 
 We multiply 2 (16) for an even (odd) number case.
 We also plot $0.8 \sin (2\phi)$ (dotted line) for comparison.} 
 \label{fig:pc}
\end{figure}

Finally, let us remark that the amplitude of the current exhibits a
nontrivial dependence on $\Phi$ when $N \to \infty$ with a fixed value
of $\delta N$. 
If we divide $I_{N}^{\delta N}(\phi)$ by $N$ and take the limit of $N
\to \infty$, we then have~\footnote{In this case, 
Eq.(\ref{eq:odd-N-finite}) can be neglected because the correction to
Eq.(\ref{eq:pc-limit}) is given by ${\cal O}(1/N^2)$. 
Therefore, the following argument can be applied regardless of an even
and odd number of $N$.}
\begin{equation}
 \lim_{N \to \infty} \frac{I_{N}^{\delta N}(\phi)}{N}
  = \begin{cases} 
    \displaystyle - \frac{2}{\pi^2} \sum_{n = 1}^\infty
     \frac{\sin(2 n\phi)}{n}
     \frac{1}{1-4 n^2 \delta N^2} & \text{for $\delta N = {\rm odd}$}, \\
    \displaystyle \frac{4}{\pi^2} \sum_{n = 1}^{\infty} 
     (-1)^{n+1} \frac{\sin(n\phi)}{n}
     \frac{1}{1-n^2 \delta N^2} & \text{for $\delta N = {\rm even}$}.
  \end{cases} \label{eq:pc-limit}
\end{equation}
When $\delta N \gg 1$, we sum $n$ in the above equations and obtain
\begin{align}
 \lim_{N \to \infty} \frac{I_{N}^{\delta N}(\phi)}{N} 
  = \begin{cases}
     \displaystyle 
     \frac{(\pi-2\phi) (\pi^2 -(2\phi -\pi)^2 )}{24\pi^2 \delta N^2}
     & \text{ for $0 \le \phi \le \pi$ and $\delta N = {\rm odd}$,} \\ 
     \displaystyle - \frac{\phi \left( \pi^2 - \phi^2 \right)}{3\pi^2 \delta N^2}
     & \text{ for $-\pi \le \phi \le \pi$ and $\delta N = {\rm even}$,}
    \end{cases} \label{eq:limit}
\end{align}
where we have used the following mathematical formula:
\begin{align}
 \sum_{n=1}^\infty \frac{\sin (n\phi)}{n^3} = 
 \frac{(\pi-\phi)\left( \pi^2 - (\phi-\pi)^2 \right)}{12} \
 & \text{ for $0 \le \phi \le 2\pi$}, \\
 \sum_{n=1}^\infty (-1)^{n+1} \frac{\sin (n\phi)}{n^3} = 
 \frac{\phi \left( \pi^2 - \phi^2 \right)}{12} \
 & \text{ for $-\pi \le \phi \le \pi$}.
\end{align}
In Fig.~\ref{fig:pc}, we plot the persistent currents of
Eq.(\ref{eq:limit}) for an even and odd number of $\delta N$ as a
function of $\phi$.
It should be noted that the persistent currents are not standard
saw-tooth curves even though we are considering non-interacting
electrons, and the functional shape of Eq.(\ref{eq:limit}) for $\delta
N$ is an odd number similar to $\sin (2\phi)$ but not identical (see
Fig.~\ref{fig:pc}). 
The solid line ($\delta N =1 = \text{odd}$) in Fig.~\ref{fig:g-ene-dN}
may be regarded as an (approximate) example of Eq.(\ref{eq:pc-limit})
and shows a smooth curve as a function of $\Phi$. 

Summarizing our results~\footnote{These results are valid when $Q$ is a
multiple of $N$.}:
(1) If $N$ is an even number, there exists exact one-half flux
periodicity when $\delta N$ is an odd number; 
(2) In the long length ($Q \gg 1$) and large diameter ($N \gg 1$) limit,
a fundamental flux period of the ground-state becomes fractional as
$\Phi_0/Z$, depending on the ratio of twist $\delta N$ to $N$
($1/Z=\delta N/N$), although we do not assume any interaction that can
form a quasiparticle of charge $Z e$; and 
(3) As $N \to \infty$ with fixed value $\delta N$, the currents are not
standard saw-tooth as is normally expected from non-interacting theories
at zero temperature.

The primary factor in these results is the Fermi surface structure of
the square lattice.
It appears that many energy bands cross the Fermi level when we roll a
planar sheet into a cylinder, and electrons near the Fermi level in each
energy band contribute to the persistent currents which interfere with
one another due to the twist.  
This point can be made clear by comparing a square lattice with a
honeycomb lattice (which possesses only two distinct Fermi points).
In this case, we can observe at most a $\Phi_0/2$
periodicity~\cite{Sasaki}.

Finally we briefly discuss the dependence of persistent currents on the
value of $Q$.
So far we implicitly assumed that $Q$ is a multiple of $N$. 
For a general $Q$ and a general structure (different chiral and
translational vectors), the persistent currents are not as
simple~\cite{Sasaki} as the results obtained up to now.
For example, when the remainder of $Q/N$ is an odd number, the one-half
periodicity may emerge even in the absence of twist (see
Fig.~\ref{fig:g-ene-Q181}). 
Furthermore, another fractional period may appear for a specific value
of $Q$ in a certain limit. 
This complication can be understood by Eqs.(\ref{eq:bri-1}) and
(\ref{eq:bri-2}). 
For $\mu_1$-th energy band, we have the occupied states
\begin{equation}
 \left[ -Q \left( \frac{1}{2} - \frac{|\mu_1|}{N} \right) 
  + \frac{\delta N}{N} \mu_1 + N_\Phi \right] + 1
 \le \mu_2 \le 
 \left[ Q \left( \frac{1}{2} - \frac{|\mu_1|}{N} \right) 
  + \frac{\delta N}{N} \mu_1 + N_\Phi \right].
 \label{eq:region-twist}
\end{equation}
Suppose $Q$ is an even number and we set $Q = n N + \delta Q$ ($n$ is an
integer).
Then the above inequality reduces to 
\begin{align}
 -\left( \frac{Q}{2} - n |\mu_1| \right) + & \left[ \frac{\mu_1 \delta N
 + |\mu_1| \delta Q}{N} + N_\Phi \right] + 1
 \le \mu_2 \nn \\
 & \le 
 \left( \frac{Q}{2} - n |\mu_1| \right) + \left[ \frac{\mu_1 \delta N-
 |\mu_1| \delta Q}{N} + N_\Phi \right],
\end{align}
which shows an asymmetry of the left and right Fermi point in the
$\mu_1$-th energy band because of the $(\delta Q/N) |\mu_1|$ term.
In fact, this term seems not to be regarded as the gauge field presented
in this paper.
However, if $\delta N = 0$ we can obtain several consequences from the
point of view of twist-induced gauge field.
We consider both $\mu_1$-th energy band in the first (second) quadrant
and $-\mu_1$-th energy band in the third (fourth) quadrant of the
Brillouin zone (see Fig.~\ref{fig:Fermi-line}), and examine the
persistent current contributed from those regions.
In this case, for a fixed value of $\mu_1$, we obtain the following
region of $\mu_2$, in which all states are occupied by electrons: 
\begin{align}
 -\left( \frac{Q}{2} - n |\mu_1| \right) + \left[ -\frac{\delta Q}{N}
 \mu_1 + N_\Phi \right] + 1 \le \mu_2 \le 
 \left( \frac{Q}{2} - n |\mu_1| \right) + \left[ -\frac{\delta Q}{N}
 \mu_1 + N_\Phi \right].
 \label{eq:region-Q}
\end{align}
This shows that the remainder $\delta Q$ can be regarded as a
twist-induced gauge field and have a similar effect as the twist, which
is manifest when we compare Eq.(\ref{eq:region-Q}) with
Eq.(\ref{eq:region-twist}). 
Therefore, we can apply the same analysis to a non-vanishing remainder
of $Q/N$ even in the absence of twist.

\section{Discussion and summary}\label{sec:sum-dis}

Here, we discuss some possible extensions of our results.
Because we have observed a fractional periodicity (in a certain limit)
in the ground state, it may be interesting to ask ``Is it possible that
the persistent currents exhibit multiple periods such as $2\Phi_0$ or
$3\Phi_0$ as a fundamental period?''
To answer this question, we consider higher genus materials ($g$: number
of holes), the ground state of which exhibit $g\Phi_0$ periodicity
depending on the genus $(g = 2,3,\cdots)$~\cite{SKS-genus}.
Moreover, there are other examples where the charge of a quasiparticle
itself becomes fractional.
The quasiparticle in the quantum Hall effect has fractional charge
$e/3$~\cite{QHE} and there is a model containing a fractional $e/2$
charged soliton in 1+1 dimensions~\cite{JR}.
Those systems might exhibit $3\Phi_0$ or $2\Phi_0$ periodicity in the
ground state, respectively.
Considering the possibility that the system exhibits flux periodicity
such as $(2/3)\Phi_0, (2/5) \Phi_0$, when we consider a torus with $g=1$
such a periodicity seems impossible.
However, for higher genus materials ($g \ge 2$), there is still a chance
for a fundamental flux period of $(g/Z)\Phi_0$, where $Z$ is an integer.
Such a periodicity may exist when the twist and genus can be
simultaneously defined for a higher genus material, although we have no 
idea if one can define twist in higher genus materials.

\begin{figure}[htbp]
 \begin{center}
  \psfrag{a}{(a)}
  \psfrag{b}{(b)}
  \psfrag{D}{$\delta N$}
  \psfrag{N}{$N$}
  \psfrag{dN}{$\bar{N}$}
  \psfrag{dM}{$\bar{M}$}
  \psfrag{p}{$\Phi$}
  \includegraphics[scale=0.4]{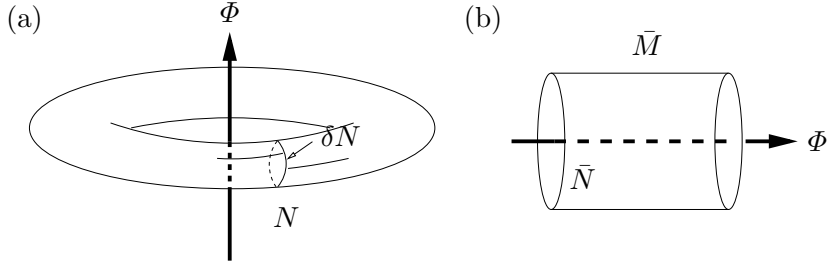}
 \end{center}
 \caption{Schematic diagram of (a) a twisted torus studied in this paper
 and (b) a cylindrical geometry studied by Cheung et al.~\cite{Cheung}. 
 Both surfaces are made from a square lattice.} 
 \label{fig:cylin-torus}
\end{figure}

We would like to mention the relationship between our results and the
previously published literature on a nontrivial flux periodicity of
persistent current.
Cheung et al.~\cite{Cheung} examined persistent currents in a finite
height cylinder whose surface is made from two-dimensional square
lattice.
They found that when the cylinder is threaded by a magnetic flux, the
persistent current is approximated by
\begin{align}
 \frac{I_0}{\pi} \sum_{n=1}^\infty \frac{\sin (n\phi)}{n} (1+ \cos ( n
 \bar{N} \pi )) \frac{\sin\left( \frac{\pi}{\bar{M}+1} \right)}{\cos
 \left( \frac{n \bar{N} \pi}{\bar{M}+1} \right) - \cos \left(
 \frac{\pi}{\bar{M}+1} \right)},
 \label{eq:cheung}
\end{align}
where $I_0$ is a constant amplitude and $\bar{M}$ is the number of
lattice site in the height direction and $\bar{N}$ in the circumference
(see Fig.~\ref{fig:cylin-torus}(b)).
They numerically checked for several samples that flux periodicity
becomes (approximately) a fraction depending on the aspect ratio of
height and circumference.
It is valuable to note that Eq.(\ref{eq:cheung}) becomes almost
identical to Eq.(\ref{eq:pc-func-t}) if we replace $\bar{N} \to \delta
N$ and $2(\bar{M}+1) = N$. 
Thus, fractional flux periodicity can be verified when
$\bar{N}/2(\bar{M}+1) = 1/Z$ in the large height $\bar{M} \gg 1$ limit.
The same configuration with a magnetic field applied perpendicular to
the cylindrical surface was analyzed by Choi and Yi~\cite{Choi}.
They reported that a fractional flux periodicity appears in the
persistent currents depending on the perpendicular magnetic field and
the number of lattice sites along the circumference.
We also note that $\Phi_0/2$ periodicity is observed in the
conductivity~\cite{AAS} (not persistent currents) for cylindrical
geometries.

In this paper, we presented an example in which the fundamental flux
period of the ground-state does not coincide with the flux periodicity
of the constituent particle.
We showed that a fractional flux periodicity
($\Phi_0/3,\Phi_0/4,\cdots$) can be realized in the long length ($Q \gg
1$) and large diameter ($N \gg 1$) limit of twisted tori made of square
lattice for half-filling cases, and the correction to the amplitude of
persistent currents scales as ${\cal O}(N/Q^3)$.
As for one-half flux periodicity ($\Phi_0/2$), we found that if $N$ is
an even number, an odd number of $\delta N$ or $\delta Q$ gives an exact 
$\Phi_0/2$ periodicity in the persistent currents.
Table~\ref{tab:1} summarizes the relationship between the lattice
structure and flux periodicity of persistent currents, derived in this
study. 
First, we classify the tori into two types depending on whether or not
$\delta Q$ is zero, where $\delta Q$ is defined as the remainder of
$Q/N$.
As we have discussed in the text, for $\delta Q \ne 0$ and $\delta N=0$,
we can obtain the similar results for $\delta Q = 0$ by setting $\delta
N = \delta Q$.
Furthermore, for the case of $N \to \infty$ with a fixed value of
$\delta N (\gg 1)$, the persistent currents are not standard saw-tooth
curves as expected from the non-interacting theories at zero temperature.
\begin{table}
 \begin{center}
 \begin{tabular}{c|c|c}
  \hline
  \multicolumn{3}{c}{\bf Lattice Structure and Flux Periodicity}\\
  \hline\hline
  \multicolumn{1}{c|} {$\delta Q$} & {$\delta N$} & Period (in
  units of $\Phi_0$)\\
  \hline
  {$\delta Q=0$} & {$\delta N=\text{odd}$} & {1/2 for $N= \text{even}$ (exact)} \\
  \cline{2-3}
  & {$\delta N \ne 0$} & {$1/Z$ for $\delta N/N = 1/Z$ (approximate)} \\
  \cline{1-3}
  \hline
  \multicolumn{3}{c}{$\delta Q \leftrightarrow \delta N$} \\
  \hline
 \end{tabular}
 \end{center}
 \caption{Nontrivial flux periodicity of tori.
 For an even number of $N$, there exists one-half flux periodicity when
 $\delta N$ ($\delta Q = 0$) or $\delta Q$ ($\delta N =0$)  is an odd
 number.
 In the long length ($Q \gg 1$) and large diameter ($N \gg 1$) limit, we
 obtain a fractional flux periodicity depending on $\delta N/N$ or
 $\delta Q/N$.} 
 \label{tab:1}
\end{table}

\section*{Acknowledgments}

K. S. is supported by a fellowship of the 21st Century COE Program of
the International Center of Research and Education for Materials of
Tohoku University. 
R. S. acknowledges a Grant-in-Aid (No. 13440091) from the Ministry of
Education, Japan.

\appendix
\section{Persistent current and Fermi velocity}\label{app:pc}

We prove that the amplitude of the persistent current is approximated by
Eq.(\ref{eq:pc-amp}) for a long system ($Q \gg 1$).
For simplicity, let us first consider an untwisted torus and fix $Q$ as an
even numbered multiple of $N$~\footnote{Note that throughout this
Appendix, $Q$ is fixed as a multiple of $N$.}.  
Then
\begin{equation}
 \left[-Q\left( \frac{1}{2} - \frac{|\mu_1|}{N} \right) + N_\Phi \right] = 
  -Q\left( \frac{1}{2} - \frac{|\mu_1|}{N} \right), 
  \label{eq:region}
\end{equation}
holds for $0 \le N_\Phi < 1$ because the right hand side of
Eq.(\ref{eq:region}) is an integer. 
It follows from
Eqs.(\ref{eq:eigen-1}),(\ref{eq:vac-ene-mu_1}),(\ref{eq:pc-mu_1}) and
(\ref{eq:region}) that for $0 \le N_\Phi < 1$
\begin{align}
 I(\mu_1, \Phi) 
 &= \sum_{\mu_2= -Q\left( \frac{1}{2} - \frac{|\mu_1|}{N} \right)+1}^{
 Q\left( \frac{1}{2} - \frac{|\mu_1|}{N} \right)} 
 \frac{2t}{\Phi_0} \frac{2\pi}{Q}
 \sin\left( \frac{2\pi (\mu_2 -N_\Phi)}{Q}
 \right) \nn \\
 &= \frac{2t}{\Phi_0} \frac{2\pi}{Q} \sin \left( \frac{2\pi |\mu_1|}{N} \right)
 \frac{\sin \left( \frac{\pi}{Q} (1-2 N_\Phi) \right)}{\sin \left(
 \frac{\pi}{Q} \right)} \label{eq:odd-f} \\  
 &= \frac{ev_F(\mu_1)}{|T_w|} \left( (1-2N_\Phi) + {\cal O}\left( \frac{1}{Q^2}
 \right) \right), 
\end{align}
where we have used Eq.(\ref{eq:fermi-vel-eff-mass}) in the last line.
This shows that the amplitude of the persistent current of $\mu_1$-th
energy band is approximated by Eq.(\ref{eq:pc-amp}) and a correction on
the order of $I(\mu_1){\cal O}(1/Q^2)$ can be ignored for long systems of
$Q \gg 1$.  
Similarly, for an odd number of $Q$ up to ${\cal O}(1/Q)$ we have 
\begin{equation}
 I(\mu_1, \Phi) = \frac{ev_F(\mu_1)}{|T_w|}(-2N_\Phi) \ \ 
  \text{for $-\frac{1}{2} \le N_\Phi < \frac{1}{2}$}.
\end{equation}
Here, it is important to mention the functional form of the persistent
current. 
The persistent current for the $\mu_1$-th energy band, which is valid
for any $N_\Phi$, is obtained by extending $\sin \left( \frac{\pi}{Q}
(1-2 N_\Phi) \right)$ in $I(\mu_1, \Phi)$ (Eq.(\ref{eq:odd-f})) as an
odd function of $N_\Phi$.
Therefore, it can be expanded by the Fourier series:
\begin{align}
 \sum_{n=1}^\infty C_n \sin (2\pi n N_\Phi),
\end{align}
with an appropriate Fourier coefficient $C_n$.
This fact will be useful to understand an exact one-half flux
periodicity, which is valid for an odd number of $\delta N$ ($\delta Q$)
with an even number of $N$ (see Eq.(\ref{eq:pc-func-t}) and discussion
below Eq.(\ref{eq:Coe})).

Next, we consider a twisted torus where $Q$ is an even numbered multiple
of $N$. 
Then
\begin{align}
 \left[-Q\left( \frac{1}{2} - \frac{|\mu_1|}{N} \right) + \frac{\delta
 N}{N} \mu_1 + N_\Phi \right] =   
 -Q\left( \frac{1}{2} - \frac{|\mu_1|}{N} \right), 
 \label{eq:region-2}
\end{align}
holds for $0 \le \frac{\delta N}{N} \mu_1 + N_\Phi < 1$ because the
right hand side of Eq.(\ref{eq:region-2}) is an integer. 
It follows from
Eqs.(\ref{eq:eigen-1}),(\ref{eq:vac-ene-mu_1}),(\ref{eq:pc-mu_1}) and
(\ref{eq:region-2}) that for $(\delta N/N)\mu_1^0 \le N_\Phi < 1-(\delta
N/N)\mu_1^0$
\begin{align}
 \sum_{\mu_1=\pm \mu_1^0} I(\mu_1, \Phi) 
 &= \sum_{\mu_1=\pm \mu_1^0}
 \sum_{\mu_2= -Q\left( \frac{1}{2} - \frac{|\mu_1|}{N} \right)+1}^{
 Q\left( \frac{1}{2} - \frac{|\mu_1|}{N} \right)} 
 \frac{2t}{\Phi_0} \frac{2\pi}{Q}
 \sin\left( \frac{2\pi (\mu_2 -\frac{\delta N}{N} \mu_1- N_\Phi)}{Q}
 \right) \nn \\
 &= \sum_{\mu_1=\pm \mu_1^0}
 \frac{2t}{\Phi_0} \frac{2\pi}{Q} \sin \left( \frac{2\pi \mu_1^0}{N} \right)
 \frac{\sin \left( \frac{\pi}{Q} \left( 1-2 \left( N_\Phi-\frac{\delta N}{N}
 \mu_1 \right) \right) \right)}{\sin \left( \frac{\pi}{Q} \right)} \nn \\  
 &= 2 \frac{ev_F(\mu_1^0)}{|T_w|} \left( (1-2N_\Phi) + {\cal O}\left(
 \frac{1}{Q^2} \right) \right).
\end{align}
This shows that the amplitude of the persistent current of $\mu_1$-th
energy band is approximated by Eq.(\ref{eq:pc-amp}) and a correction on
the order of $I(\mu_1){\cal O}(1/Q^2)$ can be ignored for long systems
of $Q \gg 1$.
As shown in Eqs.(\ref{eq:pc-untwist}), (\ref{eq:1/3}) and
(\ref{eq:pc-limit}) the summation of $\mu_1$ results in, roughly
speaking, the change from $\sin \left( 2\pi \mu_1/N \right)$ to $N$.
Hence, the correction to the persistent currents ($I_{\rm pc}^{\delta
N}(\phi)$) is given by ${\cal O}(N/Q^3)$.


\end{document}